# A colorful origin for the genetic code

## Information theory, statistical mechanics and the emergence of molecular codes

Tsvi Tlusty

*Department of Physics of Complex Systems,*

*Weizmann Institute of Science, Rehovot, Israel, 76100,*

*E-mail: tsvi.tlusty@weizmann.ac.il*

**Abstract**

The genetic code maps the sixty-four nucleotide triplets (codons) to twenty amino-acids. While the biochemical details of this code were unraveled long ago, its origin is still obscure. We review information-theoretic approaches to the problem of the code's origin and discuss the results of a recent work that treats the code in terms of an evolving, error-prone information channel. Our model – which utilizes the rate-distortion theory of noisy communication channels – suggests that the genetic code originated as a result of the interplay of the three conflicting evolutionary forces: the needs for diverse amino-acids, for error-tolerance and for minimal cost of resources. The description of the code as an information channel allows us to mathematically identify the fitness of the code and locate its emergence at a second-order phase transition when the mapping of codons to amino-acids becomes nonrandom. The noise in the channel brings about an error-graph, in which edges connect codons that are likely to be confused. The emergence of the code is governed by the topology of the error-graph, which determines the lowest modes of the graph-Laplacian and is related to the map coloring problem.

1. **Introduction**

In the cell, information is carried by molecules. Often, the organism needs to translate information written in one species of molecules into another molecular language. This requires a molecular code [1]. Perhaps the best known example is the genetic code that maps nucleotide triplets to amino-acids. Like any other code that relies on molecular recognition, the genetic code is prone to recognition errors. This inherent noise poses the organism with a fundamental question that is essential to its survival: *how to design a molecular code that can withstand the impact of noise while accurately and efficiently translating information?* We examine this general question in the context of the genetic code. Moreover, we argue that the evolutionary interplay between accuracy, efficiency and resistance to noise drove the emergence of the code [1-5].

The genetic code can be seen as a dictionary that translates the four-letter language of the nucleic bases, A, T, G, and C, into the twenty-letter language of the amino-acids [6]. Each of the sixty-four "words" or symbols in this dictionary is a triplet of nucleic bases called codon. The meaning of each codon is one of the twenty amino-acids that make up the proteins or a punctuation mark signaling the end of protein synthesis (Fig. 1A). Thus the design of an optimal code is a semantic challenge of wisely assigning meanings to symbols. The translation apparatus cannot discern well between T and C in the third position of the codon [7-9], and the effective number of discernable codons is therefore somewhere in the range of 48-64. Since there are at least 48 discernable codons and only 20 amino-acids, the code is highly degenerate or *redundant* in the sense that all amino-acids, except methionine and tryptophan, are encoded by multiple codons. Hence, there are many synonymous codon symbols that bear the same amino-acid meaning.

Observing the code-table, order is evident. Firstly, the synonymous codons that encode a certain amino-acid tend to clump together in a contiguous domain of the table (Fig. 1A). Secondly, codons that are adjacent in the table tend to encode similar amino-acids. For example, there is a good chance that they are of similar polarity [10-14], side-chain size and other chemical characteristics [15-20]. Therefore, if one draws a topographical map in which the horizontal coordinate is the location in the code-table and the altitude is the chemical property (Fig. 1B) then the resulting landscape will be *smooth* (a smooth mapping is one that maps close-by points in one space to close-by points in another space). Later we discuss why a planar code-table cannot actually capture in full the natural topology of the code.

Any theory for the code's origin must explain this notable order and smoothness. A hint may come from the discovery that the standard genetic code is not strictly universal. A few dozens of bacteria, protozoa and mitochondria use variants of the standard code [8, 9]. This suggests that the code was open, at least for some time, to evolutionary change, although it is not clear whether any of these variants offer an advantage over the standard code [20]. All the variants of the code use the same twenty amino-acids, which raises the question whether this number has any special significance.

Many theories have been put forward to answer the questions about the origin of the smooth genetic code with its twenty amino-acids. One answer that came from Crick and others, is that this is merely a 'frozen accident' [6, 21], that the twenty amino-acids and their assignment to codons were good enough to work and this number is too rigid to change, since any further alteration of the code would imply a devastating change of nearly all the proteins (although Crick also suggests that the code could become smooth before it froze). A second possibility is that the current shape of the code and the number twenty depend finely on the specifics of primordial biochemical interactions [22-30] and pathways [31-34] or on the values of evolutionary parameters [35-39] (for surveys of theories for the code's origin see [21, 36, 40, 41]). Here we explore a third possibility, that the pattern and number of amino-acids are fundamental topological features of the noisy information channel that is embodied in the genetic code. We hypothesize a minimal generic model for code evolution, and suggest that the number of amino–acids may arise from the fact that the code is an assignment of amino-acids to triplets of a four-letter alphabet.

A body of theoretical works proposes that the order in the code can be explained by the evolutionary selection for codes that minimize the deleterious impact of translation errors and mutations, termed *error-load*: A mistranslation, or a mutation, of one base in a codon should lead to minimal change in the chemical nature of the translated amino-acid. The need for resilience to errors is an evolutionary force that drives adjacent amino-acids to be similar as possible, resulting in a *smooth* code. Due to the smoothness of the code, there is fair chance that a single-base error would yield the same amino-acid – and even if it does not – it is likely to produce a chemically similar amino-acid (Fig. 1). However, to synthesize functional and efficient proteins the code-table must encode a diverse enough catalogue of amino-acids. Thus *diversity* is a counteracting evolutionary force, which drives the code to be as heterogeneous as possible. The need for diversity may be the force that drives certain organisms to expand their amino-acids catalogue to non-canonical amino-acids [42, 43]. Since the genetic code is realized in molecules, it costs the organism materials, energy and time to synthesize the molecules, archive their blueprints as genes, replicate the genes and correct any mutational damage. Therefore, the third evolutionary force that we should consider is this *cost* of the coding system.

The hypothesis that the evolution of the code was governed by the need to minimize the error-load was brought on almost simultaneously with the deciphering of the code itself [10-13, 44-46]. Quantitative models verified that, indeed, codons which are likely to be confused tend to encode the same or similar amino acids and that it is practically impossible to obtain such a smooth pattern by random assignment [18-20, 47-57]. Evolutionary dynamics that could drive the genetic code towards such optimal error-resilience was demonstrated in a simplified simulation model in which the organisms that utilized a code with superior error-resilience took over the population [35]. In a series of studies, Sella and Ardell [21, 37, 38] constructed more realistic simulation models that incorporated into the dynamics the essential co-evolution of the coding machinery, the nucleotide sequence it reads and the proteins into which the sequence is translated [58, 59]. Their models were capable of producing genetic codes that are similar to the naturally genetic code in their redundancy and smoothness. However, the codes in their simulations often froze at a suboptimal configuration, whereas the standard genetic code is known to be highly optimized. A recently suggested explanation, which was supported by a simulation study [60], is that the highly optimized genetic code is the outcome of recurrent horizontal gene transfer.

Motivated by the evidence that supports the error-load minimization hypothesis, we constructed a rigorous framework in order to examine the emergence and evolution of molecular codes [1-5], in particular the genetic code, in terms of rate-distortion theory [61-63]. The present paper reviews recent results of this rate-distortion approach, focusing on its basic concepts and leaving aside the exact details of the formalism. In what follows, we briefly depict the primordial circumstances which could lead to the emergence of the code. Then, we describe the genetic code as an information channel that maps the symbol space (the codons) into the space of meanings (the amino-acids). This enables us to calculate the fitness of the code, which takes into account the interplay between the error-load, the diversity and the cost of resources which the organism needs to invest in the construction of the channel (Section 2). The biological challenge of maximizing the fitness of the genetic codes is shown to be equivalent to the communication engineering problem of designing an optimal information channel. The noise in the channel delineates the structure of the codon-space, which is a graph defined by the most probable reading errors (Section 3). Next, we examine a population of organisms that compete according to the fitness of their codes. The model suggests a generic mechanism for the emergence of a genetic code as a phase transition in the information channel, where the lowest excited modes of the codon-graph correspond to the encoded amino-acids (Section 4). This phase transition suggests that the topology of the codon-space limits the number of the lowest modes, thus the number of amino-acids (Section 5). We then summarize the predictions of the present model and its underlying assumptions (Section 6). Finally, we briefly discuss the application of the present framework to other molecular codes (Section 7).

2. **The genetic code as a noisy information channel**

The genetic code is a mapping, or an information channel, that receives an input of codon symbols and outputs their corresponding amino-acid meanings [1-5] (Fig. 2A). Our scenario imagines primitive cells in which amino-acids and nucleotides coexist and bind *non-specifically* [44, 64-69]. Even such random association may assist the primitive living systems in catalyzing their metabolism and self-reproduction: Some suggest that non-specific binding of nucleotides could catalyze the synthesis of random polypeptides, barely functional proto-proteins [13, 45, 59, 60, 64, 65, 70-72]. Others propose that amino-acids catalyze nucleotide synthesis in the RNA-world [67, 73]. Of course, *specific* binding of nucleotides to amino-acids would benefit the organism much more by enabling controlled synthesis of programmed, functional proteins. However, this specificity comes at a cost of producing and maintaining molecular recognizers, such as tRNA, that could discern their target among similar lookalikes. As we discuss below, a code will emerge exactly at the point when the benefit of specificity surpasses its cost.

In the present model, we start from the state of ambiguous, random association of amino-acids and nucleotides [13, 21, 37, 38, 44, 45, 59, 60, 69] termed the *'non-coding'* state. The ambiguous starting point disregards certain idiosyncratic determinants that may have influenced the early evolution of the code, such as stereo-chemical affinity and the history of biochemical pathways. Nevertheless, this enables the model to examine in an unbiased manner the hypothesis that the genetic code appeared at a coding transition in an information channel. We suggest that this transition is governed mainly by the cost and quality of this information channel and less by the biochemical details of the association of amino-acids and nucleotides.

To formalize the balance of evolutionary forces that leads to the emergence of a code, we consider an information channel that relates the space of $N_A$ amino-acids "meanings", $\alpha, \beta, \gamma\ldots$, and the space of $N_C$ codon "symbols", $i, j, k\ldots$ (Fig. 2A) [1-5]. To define the channel one writes down the probabilities $p_{i\alpha}$ that a codon $i$ encodes an amino-acid $\alpha$. These $N_C \times N_A$ probabilities form a *code-matrix* [61, 62]. Initially, the correspondence between amino-acids and codons is completely non-specific. Any codon is equally likely to encode any of the amino-acids and the code-matrix is therefore uniform, $p_{i\alpha} = 1/N_A$. At this state, the channel conveys no information because there is no correlation between the input codon and the randomly drawn amino-acid at the output. This state is appropriately termed *non-coding*. A code emerges only when at least some of the codons tend to encode preferred amino-acids and the code-matrix thereby becomes non-uniform.

An instructive way to visualize the code matrix is by assigning to each of the $N_C$ codons an arrow whose direction represents the probability of this codon to encode any of the $N_A$ available amino-acids (Fig. 2B). Thus it is evident that the coding system is analogous to a system of magnetic spins, an analogy that will be used in discussing the coding transition (Section 4). Each codon corresponds to a magnetic "spin" and each amino-acid corresponds to one of the axes of an $N_A$-dimensional space in which the spin resides. The spin of codon $i$ is the vector made of the $i$-th row of the code matrix ($p_{i\alpha}$, $p_{i\beta}$, $p_{i\gamma}$ …). The code matrix $p_{i\alpha}$ is simply the probability that the $i$-th spin is oriented in the $\alpha$-th direction. Spins pointing in a certain direction are analogous to codons that tend to encode a certain amino-acids.

The fitness of the genetic codes consists of contributions of three evolutionary forces: the error-load, the diversity and the cost. The *error-load* depends on the accuracy of reading. Since the molecular reading apparatus is imperfect it may sometimes confuse between two codons. The probabilities for all such misreading events between the $N_C$ codons form the $N_C \times N_C$ reading matrix $r_{ij}$. The diagonal terms $r_{ii}$ are the probabilities to correctly read the codons, whereas the non-diagonal terms are the misreading probabilities. Since misreading may result in the wrong amino-acid, we also need a measure for chemical distances among amino-acids, which is expressed in terms of a distance matrix $c_{\alpha\beta}$ (whose diagonal elements vanish $c_{\alpha\alpha} = 0$). This distance should account for chemical characteristics like polarity [10-14], size and hydrophobicity [15-20]. The overall error-load is the average chemical distance $c_{\alpha\beta}$ between all possible pairs of encoded amino-acids, $\alpha$ and $\beta$, at all pairs of codons, $i$ and $j$, that are likely to be confused by misreading, weighed by the probability that such a misreading occurs, $r_{ij}$,

$$\text{(1)} \quad \text{error-load} = \sum_{i,j,\alpha,\beta} r_{ij} \, p_{i\alpha} p_{j\beta} c_{\alpha\beta}.$$

The error-load measures the *smoothness* of the code. In a smooth code nearby codons that are likely to be confused encode similar amino-acids, thus reducing the impact of errors. A smooth code therefore has a lesser error-load than a "rugged" one (Fig 1B).

An evolutionary force that counteracts the error-load is the need for *diversity*. The versatility and efficiency of the proteins that the organism synthesizes benefit from the diversity of the available amino-acid building blocks. This fitness benefit is estimated by the average chemical distance between all the encoded amino-acids,

(2) $$\mathtt{diversity} = \sum_{i,j,\alpha,\beta} \left(1-\delta_{ij}\right) p_{i\alpha} p_{j\beta} c_{\alpha\beta}.$$

Clearly, if all the codons encode the same amino-acid then the diversity vanishes, whereas the diversity is maximal if the codons encode the most distant amino-acids. It is evident that while the error-load depends on the "geometry" of the symbol-space through the reading matrix $r_{ij}$, the diversity is geometry-independent. By varying the misreading matrix, one may therefore vary the error-load while keeping the diversity unchanged.

Molecular codes relate symbols and meanings via biochemical recognition and binding. The code-matrix $p_{i\alpha}$ is simply the probability that the molecule carrying the meaning (amino-acid) $\alpha$ binds the molecular symbol (codon) $i$. The *cost* of molecular codes is mostly due to the need to ensure the specificity of this binding. High specificity increases the chance that the molecular recognizers will bind to their designated targets and thus reduce the chance of misreading. We therefore estimate that the cost of the code is proportional to the average binding energy [1, 3, 4]. Since, due to the Boltzmann distribution, the binding energy scales like $\sim \ln p_{i\alpha}$, we find that the cost is

(3) $$\mathtt{cost} = \sum_{i,\alpha} p_{i\alpha} \ln\left(\frac{p_{i\alpha}}{p_\alpha}\right),$$

where $p_{i\alpha}$ is normalized by the sum $p_\alpha = (1/N_C) \sum_j p_{j\alpha}$. The normalization ensures that the cost at the non-coding state vanishes because the codons and the amino-acids are uncorrelated and the code-matrix is therefore $p_{i\alpha} = p_\alpha = 1/N_A$. As specificity increases the cost increases towards its upper bound, $N_C \ln N_C$, which occurs when every codon encodes exactly one amino-acid.

Optimizing the code amounts to balancing the error-load, the diversity, and the cost. It proves convenient to describe this balance as the maximization of the overall code fitness,

(4)    `fitness= − error-load + `$w_D$`×diversity − `$w_C$`×cost.`

The diversity parameter $w_D$ weighs the significance of diversity in the overall fitness, whereas the cost weight $w_C$ measures the evolutionary burden of the resources required by the code. Large $w_C$ drives the code to reduce its specificity because the price of the binding sites rises, whereas increasing $w_D$ drives the code to higher diversity and specificity. The cost and the error-load are prefixed with a minus sign because they reduce the fitness, while the diversity increases the fitness. It is evident from Eqs.(1-4) that the optimal code, the code-matrix $p_{i\alpha}$ which maximizes the fitness, is a function of four determinants: the reading matrix $r_{ij}$, the distance matrix $c_{\alpha\beta}$, and the two weights, $w_C$ and $w_D$.

Below we discuss how the genetic code evolves in response to changes in these parameters. In particular, the emergence of the genetic code appears as a phase transition in the noisy information channel described by the code matrix. Finally, we note that analogous expressions are used in communication engineering to optimize a noisy channel. A combination of the error-load and the diversity measures the average *distortion* of a message passing through the channel, whereas the cost is in fact the *rate* of information flow [1, 61-63]. The channel is optimized by minimizing a linear combination of the rate and the distortion which is equivalent to the fitness (Eq. 4).

### 3. **The topology of the genetic code**

Before we discuss the emergence of the code, we need to consider the structure and the topology of the spaces in which it emerges, the amino-acid and the codon spaces. The geometry of the amino-acid space is determined by the chemical distance, $c_{\alpha\beta}$. In the codon space, the reading and misreading probabilities provide us with a geometric notion of proximity and distance [2]. Two codons $i$ and $j$ are "adjacent" if they are likely to be confused, $r_{ij} \neq 0$. One can describe this geometry of the codon-space in terms of a graph, in which the vertices are the codons and edges connect adjacent codons (Fig. 3). In the case of the genetic code, it is plausible to assume that two codons are linked by an edge if all their letters except for one agree (Hamming distance = 1). This is because the majority of the errors, such as translation errors or mutations, are likely to involve a single letter difference. Mistranslations are expected to be significant in the early history of life, when the translation apparatus was rather inaccurate, and are therefore relevant to our scenario for the emergence of the code. On the graph, such single-letter errors correspond to a shift along an edge to a neighboring vertex.

Besides the traditional code-table (Fig. 1A), quite a few ingenious schemes were devised in order to draw the genetic code on a sheet of paper [74-76] or on a sphere [77] in a manner that would reflect the interrelations between codons and amino-acids. Nonetheless, as we discuss below, the natural topology of the genetic code is inherently not planar or spherical and it is very hard to justly visualize the code in these geometries [2]. To illustrate the topology of the genetic code we consider first a simpler code: A doublet code of a three-letter alphabet (Fig. 3A). The codons in this space are the nine two-letter words that one can compose of this alphabet. Each codon is connected to four adjacent codons differing by one letter [2]. The resulting graph exhibits the shape of a torus. The topology of a graph is defined by its genus $\gamma$, the minimal number of holes in a surface on which one can draw the graph such that no two edges cross [78]. In this simple example the genus is $\gamma = 1$ (a doughnut has one hole). The genus of a graph increases with the number its codons and their connectivity. One may therefore expect that the highly interconnected, 48- (or 64-) codon graph of the genetic code will be of a high genus since many holes will be required to avoid crossings. The genus of the simple toroidal code, $\gamma = 1$, is apparent by inspection. Calculating the genus of the genetic code, however, requires the application Euler's formula $\gamma = 1 - ½(V - E + F)$, where $V$ is the number of vertices of the graph, $E$ – the number of its edges and $F$ is the number of faces of the surface on which the graph is drawn [78]. The toroidal code, for example, has $V = 9$ vertices, $E = 18$ edges and $F = 9$ quadrilateral faces. The resulting genus, according to Euler's formula, is $\gamma = 1 - ½(9 - 18 + 9) = 1$. By the same counting procedure, one finds that the genus of the genetic code graph is in the range of $\gamma = 25\text{-}41$ (Fig. 3B), depending on the effective number of codons, which is somewhere between 48 and 64 [2]. This means that drawing the genetic code graph such that no edges cross would require a holey surface of at least 25 holes.

4. **The emergence of the code as a transition in a noisy channel**

On the holey surface representing the natural genetic code, neighboring codons are those that are likely to be confused by misreading. This surface is therefore the natural geometric framework to approach the evolutionary dynamics of a code that is driven to minimize the effects of misreading errors. To explore the generic features of this dynamics, we hypothesize a primordial world where primitive organisms compete by the fitness of their codes [1-5]. Every organism is specified by its code-matrix $p_{i\alpha}$ and our simple model assumes that the population of organisms peaks around an optimal value of the code (generalizations that include the effect of mutations, genetic drift and rugged evolutionary landscape are discussed elsewhere [1, 2, 4]). We assume that the code evolves in an environment where many amino-acids are available for incorporation [79, 80], and inquire why only twenty were chosen.

The optimal code is found by maximizing the fitness (Eq. 4) with respect to the code matrix $p_{i\alpha}$. In the limit of large cost parameter, $w_C$, specificity is too costly and the optimal code-matrix is uniform, i.e., in the non-coding state (Fig. 4A). When $w_C$ is reduced below a certain *critical* value, or equivalently $w_D$ is increased above a corresponding critical value, there appears a tendency of some codons to encode certain amino-acids more than the others (these critical values are derived in [1-4]). This signifies the emergence of the genetic code with correlated codons and amino-acids [3]. Close to this *coding transition*, the emergent code tends be smoother with fewer amino-acids. When the cost parameter $w_C$ is further reduced, the code tends to include the maximal number of amino-acids because the effects of errors are relatively small. The increase in $w_D$ that could lead to the emergence of the code may be a result of changes in the environment and in the organism. For example, if the environment becomes more complex, it transmits a higher information rate to the molecular sensors of the organism. Now the organism benefits more from investing resources in the diversity of the code, thus increasing $w_D$. Similarly, increasing the complexity of the organism's biochemical circuitry is expected to increase $w_D$, which is equivalent to effectively reducing $w_C$ [1, 3, 4].

The coding transition in the channel is analogous to a second-order phase transition in a statistical-mechanics ensemble of magnetic spins (e.g., [3, 81]). As discussed above (Section 2, Fig. 2B), each of the $N_C$ codons corresponds to a magnetic "spin", which may take one out of $N_A$ possible "states" given by the amino-acids. The ferromagnetic interaction induces order by orienting neighboring spins in parallel, just like the error-load (Eq.1) drives neighboring codons to encode the same or similar amino-acids. The diversity, on the other hand, drives codons to encode the most distant amino-acids. Hence diversity is an "anti-ferromagnetic" interaction in the sense that it drives the spins to point in opposing directions. Furthermore, thermal fluctuations tend to destroy the order in the magnetic system by randomizing the orientation of spins just like the cost (Eq. 3) tends to randomize the coding system. Finally, the fitness that combines the three evolutionary forces is mathematically equivalent to the free energy of the spin system [3, 4].

The order parameter which measures the degree of order in the magnetic system is the average orientation of the spins. The analogous order parameter in the coding system is the tendency of a given codon towards certain amino-acids, which is measured by the deviation $\delta p_{i\alpha}$ of the code matrix from the non-coding state, $\delta p_{i\alpha} = p_{i\alpha} - 1/N_A$. At high temperatures, thermal fluctuations overcome the magnetic interaction and the system is completely disordered, i.e. spins are randomly oriented. As the spin system is cooled down it reaches a certain critical temperature where the magnetic forces are strong enough to align neighboring spins in the same direction. The transition is termed continuous or second order because the degree of order continuously increases from zero as the temperature continues to decrease below the critical temperature. This is in complete analogy to the emergence of a genetic code at a critical coding transition where the coding order parameter $\delta p_{i\alpha}$ continuously increases from zero (Fig. 4A). The parameter that controls the coding transition, and is equivalent to the temperature in the spin system, is a combination of the diversity parameter $w_D$ and the cost weight $w_C$ [3].

Another useful physical analogy of the code is to an elastic system. The error-load involves the chemical distances between amino-acids encoded at adjacent codons. This mathematical form of the error-load is equivalent to the elastic energy of a spring framework in the shape of the codon graph (Fig. 4B). Every edge of the codon-graph corresponds to a spring in this framework. If the codons connected by such a "spring" encode chemically different amino-acids the spring is stretched and costs elastic energy, i.e. error-load. A "spring" connecting codons that encode the same amino-acid is loose and costs no error-load [3]. At this point, it is worth drawing the physical analogy of a beating drum. A drum emits a sound spectrum that is a series of modes with quantized frequencies and elastic energies determined by its shape and topology [82]. The low frequency modes of the drum are those of low elastic energy and are therefore easier to excite. Similarly, our model suggests that the dynamics of code evolution are affected by the topology of the codon graph. Close to the coding transition, only the modes with the lowest error-load will be excited. In both cases, of the beating drum and of the genetic code, the mathematical operator that describes the dynamics is the Laplacian, which is essentially the reading matrix $r_{ij}$ [1-5]. As we lack knowledge of the reading and misreading probabilities $r_{ij}$ at the time when the genetic code emerges, we cannot expect to exactly optimize the fitness of the code to find the excited modes. Instead, we pursue the properties of the modes that depend solely on the topology of the codon graph.

5. **The genetic code and the coloring problem**

The codon-graph can have several first excited modes, that is modes with the same (or similar) maximal fitness. Each of these probability modes corresponds to an amino-acid and the number of modes is therefore the number of amino-acids. Each of the first excited modes has a single maximum [83, 84] and determines a single contiguous domain where a certain amino-acid is encoded [1-5]. Indeed, all amino acids are encoded by synonymous codons arranged in contiguous domains except serine that splits into two domains (Fig. 1A). In addition, our model concludes that the genetic mode is smooth because the lowest-excited modes at the transition are the smoothest non-uniform modes.

The low modes partition the codon graph into domains. The aggregation of synonymous codons into contiguous domains minimizes the error-load by minimizing the boundaries between these domains thus maximizing the chance that misreading will result in the same amino-acid and therefore in vanishing error-load [85]. The partition into amino-acid domains may be likened to drawing borders between countries on a geopolitical map. We find that the problem of maximizing the fitness of the code by optimizing this partition is related to another classical partition problem, the coloring problem [2, 5]. The goal of the coloring problem is to calculate the minimal number of colors required to color an arbitrary map on a surface such that no two bordering countries have the same color. This minimal number is known as the *coloring number*, or the chromatic number $chr(\gamma)$, which is determined by the genus of the graph according to Heawood's formula [86]

(5) $$chr(\gamma) = \text{int}\left[\tfrac{1}{2}\left(7 + \sqrt{1+48\gamma}\right)\right]$$

where *int* denotes integer value. For example, the coloring number of a sphere ($\gamma = 0$) is $chr(0) = 4$, known as the four-color theorem.

A major finding of our hypothesized model is that the topology of the code sets an upper limit to the number of first excited modes (of the graph Laplacian), and thus to the number of amino-acids [2, 5]. This limit is the coloring number $chr(\gamma)$ given by Eq. 5. The relation of the coloring problem to the maximal number of first excited modes has a geometrical origin that may be clarified by constructing the following mapping of the codon graph (Fig. 4B): Each codon $i$ is mapped to a vector $\mathbf{w}(i)$ and each coordinate of this vector corresponds to one of the first-excited modes, i.e., $\mathbf{w}(i) = (p_{i\alpha}, p_{i\beta}, p_{i\gamma}\ldots)$. The dimension of the space into which the codon-graph is mapped (the length of the vector) is therefore equal to the number of modes. Since every mode peaks at a contiguous domain, the resulting surface has only one maximum and one minimum in any given direction [83, 84]. The dimension of surfaces that exhibit this property is known to be bounded by $chr(\gamma) - 1$ [87, 88]. The ground-state mode adds another available mode, and the total number of modes, which is the number of amino-acids, is the coloring number $chr(\gamma)$. For the natural genetic code with its 48-64 triplets of four letters the genus is in the range of $\gamma = 25\text{-}41$ and by Eq. 5 the coloring limit is in the range of $chr(\gamma) = 20\text{-}25$, in the neighborhood of the naturally occurring number [2, 5].

## 6. Assumptions, predictions and open questions

At this point, it is instructive to clarify the assumptions underlying the present rate-distortion theory and the predictions it makes. Given the codons and the probable errors that confuse these codons, one constructs an error-graph in which codons are vertices and edges are drawn between confused codons. The topology of the error-graph sets an upper limit on the number of amino-acids, namely the coloring number. This is the maximal number of smooth modes that emerge in the vicinity of the coding transition and each of these modes corresponds to an amino-acid. In the case of three-letter codons written in a four-letter alphabet, the assumption that the relevant errors involve codons differing by one letter defines the topology of the genetic code whose coloring number is 25. Taking into account the difficulty of the reading machinery to discern between U and C in the third letter, the topology is altered and the corresponding coloring limit is reduced to 20. Thus, the coloring limit for the genetic code is in the range of 20-25 amino-acids.

A general prediction of the model is that varying the balance between the rate and the distortion of a noisy molecular information channel induces a coding transition (Section 5). The inputs of the model are the codons and the probable errors, which define the graph, and its output or prediction is an upper limit on the expected number of amino-acids (Section 6). The predicted limit (Eq. 5) relies solely on the topology, as manifested by the graph's genus, $\gamma$ and requires no other input parameters. Altering the topology changes the resulting coloring number, and the model predicts how this limit increases with $\gamma$ (Table 1).

Our rate-distortion approach involves several assumptions and open questions which reflect the lack of experimental knowledge about the environment where the code emerged:

(i) The model assumes that there is an a priori assignment of nucleotide triplets to amino-acids [89] without detailing an explicit origin for such assignment (Section 2). It also ignores the question of why specifically these amino-acids were assigned and not others. Nevertheless, the model suggests that the assigned amino-acids are ordered in a smooth pattern according to similarity criteria, such as size and polarity.

(ii) The coloring number is merely an upper limit to the number of amino-acids. It is not clear at all why the coding machinery should exhaust this maximal number. The model, however, suggests that this is beneficial in maximizing the diversity of the code.

(iii) A central assumption of the model is that the code was shaped in the vicinity of the coding transition, which governs its smooth pattern. A related underlying assumption is that, not long after this phase transition, any further changes in the code became too costly due to the global impact on the proteins. In this respect, our model implies 'almost-frozen' dynamics in the spirit of Crick's arguments [6, 21]. It is possible that the dynamics was frozen by the rapid increase of selection forces after the code emerged. As organisms developed and their complexity increased, any alteration of the code entailed an intolerable modification of too many proteins [1, 3, 4]. Consequently, the only relevant modes are the smooth ones that emerge at the coding transition. The coloring number is a mathematical limit on the number of smooth modes. In that sense, it is a universal property of the dynamics which does not depend finely on the parameters but solely on the topology.

(iv) The wobble hypothesis suggests that the translation apparatus can effectively discern only 48 codons [7, 8] and the corresponding codon graph leads to the coloring limit of 20 amino-acids. However, if the wobble ambiguity is removed there are 64 discernible codons and the maximal number of amino-acids increases to 25. It is still unclear why evolution froze before it could improve the translation apparatus so it can discern all the 64 codons.

(v) The model assumes that the most probable translation errors at the time the code was determined involve codons differing by a single letter. However, the reading machinery might also allow certain two- and three-letter errors, which would change the topology and the resulting coloring number. Still, the model can treat any other error patterns as long as their topology is given (Table 1).

7. **Conclusions**

The present hypothesis examines the possibility that the code emerged as the result of evolutionary dynamics that is governed by the conflicting needs for error-tolerance, diversity and cost. When these three evolutionary forces representing the cost and the benefit of the code are in equipoise, the association of codons to amino-acids becomes specific and a code is born. This critical point is likened to a transition in a noisy information channel. This scenario is generic to noisy channels and does not depend finely on the missing details of the primordial biochemistry and evolutionary pathways. Minimal assumptions regarding the topology of codon space, by roughly estimating which are the most probable errors, yield basic traits of possible genetic codes, in particular their smooth encoding patterns (Fig. 1) and the maximal number of amino-acids.

Among the difficulties in this model [2], there remains the question of why the code should reach the maximal limit set by the coloring number. It is possible that the misreading probabilities $r_{ij}$ coevolved with the code to approach the maximal amino-acid diversity. Changing $r_{ij}$ could vary the topology of the codes by adding or deleting edges and vertices to the codon graph. As the reading accuracy increases, more codons become discernible and the effective number of vertices increases, which may vary the topology of the graph and its coloring number (Table 1). One may therefore imagine an evolutionary pathway for the expansion of the code, from simple codes with a few amino-acids [90], to the present-day code with its twenty amino-acids. A future advance may remove the ambiguity in the third position of the codon and thus may expand the code up to 25 amino-acids [91].

A main difficulty for the current hypothesis is the lack of data on error probabilities, due to mistranslation and mutations, and on the cost and diversity parameters at the time when the code emerged. Any experimental evidence would therefore be invaluable. One experimental direction would be to examine the evolution of organisms whose genetic code was expanded or shrunk to include more or less amino-acids than the optimal coloring number. The experimental feasibility of expanding the genetic code has been demonstrated through the last decade in E. coli, yeast, and mammalian cells, where around thirty unnatural amino-acids have been cotranslationally incorporated, one at a time, into proteins [91, 92]. In these studies the code was altered by introducing a modified tRNA together with a matching aminoacyl-tRNA synthetase that charges the tRNA with an unnatural amino-acid. Evolution is expected to drive the number of amino-acids back to the optimum, or at least a precursor of such change may be observable. To avoid drastic effects of modifying simultaneously a large number of proteins, one may keep the original tRNA and smoothly tune the expression level of the two competing tRNAs. Also, if the reassigned codon is rare [93-95], for example the UGA stop codon (amber), the impact on fitness would be minimized.

The present hypothesis suggests a generic model for the emergence and evolution of any noisy molecular code. Therefore, it may be possible to apply this approach to other, more accessible, molecular codes, such as the transcription regulation network [96, 97], and to artificially engineered biochemical codes. Indeed, similar error-resilience characteristic was identified even within the Watson-Crick pairing interactions between nucleotides [98]. A recent model [99] also uses the general framework of rate-distortion theory, but in a non-equilibrium formulation, and suggests a very similar scenario of punctuated equilibrium in the spirit of Table 1. In general, molecular recognition systems may be analyzed in terms of noisy information channels or codes as they are probabilistic association table of molecules [100-102]. These channels are often among the building blocks of molecular computing machinery, such as the RecA-induced homology recognition system [103-105]. The transcription regulatory network, as an example, associates the transcription factors with their specific DNA binding sites. It is therefore worth to consider these protein-DNA networks as noisy coding systems. Just like in the case of the genetic code, the space of DNA binding sites is represented as a graph in which vertices are the binding sites and edges connect site that are likely to be confused, i.e., by binding of the wrong transcription factor [97]. Similar topological and coloring considerations may explain the structure of the network and its size.

**Tables**

| Code | 1st position | 2nd position | 3rd position | No. of codons | $\gamma$ | Max. no. of amino-acids, chr$(\gamma)$ |
|---|---|---|---|---|---|---|
| 4-base singlets | 1 | 4 | 1 | 4 | 0 | **4** |
| 3-base doublets | 3 | 3 | 1 | 9 | 1 | **7** |
| 4-base doublets | 4 | 4 | 1 | 16 | 5 | **11** |
| 16 codons | 4 | 4 | 2 | 32 | 13 | **16** |
| 48 codons | 4 | 4 | 3 | 48 | 25 | **20** |
| 4-base triplets | 4 | 4 | 4 | 64 | 41 | **25** |

Table 1. **Topological limit to the number of amino-acids for various codes.** The codes are specified by the number of discernable letters in the three positions of the codons. The six codes are ordered according to the accuracy of the translation machinery. The number of discernable codons determines the genus ($\gamma$) and the topological limit on the maximal number of amino-acids is the coloring number, $chr(\gamma)$ (Eq. 5). As the machinery becomes more stringent the number of amino-acids increases. The present-day translation machinery with its 48-64 discernible codons (partial or no discrimination between U and C in the third position) gives rise to 20-25 amino-acids. One may imagine a future improvement that will remove the ambiguity in the third position (64 discernable codons) and is predicted to enable expansion of the code up to 25 amino-acids.

**Figure Legends**

Fig 1. **The genetic code, its order and smoothness**. **(A)** The 64 base-triplet codons are listed with the amino-acids or the stop signal (TER) they encode. The synonymous codons for a given amino-acids are grouped into contiguous domains in the table. **(B)** The polarity of the amino-acids (top) and the side-chain volume (bottom) describe smooth landscapes when plotted as the altitude of a "topographic" map on the code table (values are the amino acid polarity [18] and molecular volume [15]). Smoothness implies that close by codons are mapped to the same or similar amino-acids, i.e. amino-acids that are close-by in the amino-acid space. The smoothness of the code-table reduces the error-load since misreading is likely to replace an amino-acid by a chemically similar one. Later we discuss why a planar code-table does not capture in full the natural topology of the code.

Fig. 2. **The genetic code as a noisy information channel.** **(A)** The genetic code is a mapping that relates the space of the $N_C = 64$ codons $i, j, k...$ (left) and the space of the $N_A = 20$ amino-acids, $\alpha, \beta, \gamma...$(right). The information channel is determined by the probabilities $p_{i\alpha}$ that a codon $i$ encodes an amino-acid $\alpha$, which form the $N_C \times N_A$ *code-matrix* (middle). Initially, at the non-coding state, the amino-acids and the codons are uncorrelated and the code-matrix is therefore uniform, $p_{i\alpha} = 1/N_A$. A code emerges when at least some of the codons prefer certain amino-acids more than the others and the code-matrix is non-uniform, $p_{i\alpha} \neq 1/N_A$. The molecular reading apparatus may confuse between two codons, $i$ and $j$, at a probability $r_{ij}$. The probabilities for all such misreading events form an $N_C \times N_C$ *reading matrix* $r_{ij}$ (left). The reading matrix defines the geometry of the codon-graph, in which the vertices are the codons and edges connect codons that are likely to be confused. The chemical distances between amino-acids is expressed in terms of a distance matrix $c_{\alpha\beta}$, whose diagonal elements vanish $c_{\alpha\alpha} = 0$ (right). **(B)** The code matrix can be visualized by an ensemble of $N_C$ "spins" residing in $N_A$-dimensional space. Each codon corresponds to a "spin" and each amino-acid corresponds to one of the axes. In the present example, there are 3 available amino-acids, $\alpha, \beta, \gamma$, and the space is three-dimensional. The direction of the $i$-th spin is given by the $i$-th row of the code

matrix ($p_{i\alpha}$, $p_{i\beta}$, $p_{i\gamma}$ ...). The code matrix $p_{i\alpha}$ is therefore the probability that the *i*-th "spin" is oriented in the α-th direction. By conservation of probability, the ends of all spins must reside in the triangle $p_{i\alpha} + p_{i\beta} + p_{i\gamma} = 1$ (gray). Spins pointing in a certain direction are analogous to codons that tend to encode certain amino-acids. For example, a codon can encode a single amino-acid ($p_{i\alpha} = p_{i\beta} = 0$, $p_{i\gamma} = 1$; blue), a combination of two amino-acids ($p_{i\alpha} = p_{i\beta} = 1/2$, $p_{i\gamma} = 0$; green) or equally probably all three amino-acids ($p_{i\alpha} = p_{i\beta} = p_{i\gamma} = 1/3$; red), which represents the homogenous non-coding state (black dot).

Fig.3 **The topology of codon graphs**. **(A)** A simple codon-graph: The vertices of the graphs are all 9 possible doublets from the three-letter alphabet, A, B and C. Edges link codons differing by one letter (Hamming distance = 1). Therefore, single-letter errors correspond to a shift along an edge to a neighboring vertex. Each codon has four such neighbors. The resulting graph is embedded in a torus of genus $\gamma = 1$ and of coloring number $chr(1) = 7$ (Eq. 5). **(B)** The 64-codon graph of the genetic code: The figure depicts a codon in this graph with its 9 neighbors and 9 of its next-nearest neighbors. Each vertex is surrounded by 9 quadrilateral faces. Drawing the whole graph of the genetic code graph such that no edges cross would require a surface of $\gamma = 41$ holes. This genus is found by applying Euler's formula (text) [2]. If the effective number of codons is reduced to 48 by the third-letter wobble [7] the required number of holes decreases to 25.

Fig. 4. **The emergence of the code as a transition in a noisy channel: a geometrical view.** **(A)** The figure graphs schematically the evolution of the code as the cost and diversity parameters, $w_C$ and $w_D$, are varied (the double arrow indicates that $w_D$ increases leftwards and $w_C$ rightwards). Plotted are the cost (blue), the diversity (brown), the error-load (green) and the fitness (red). In the limit of large $w_C$ and small $w_D$, specificity is too costly and the channel is in the non-coding state of uncorrelated codons and amino-acids. When $w_C$ is decreased below a critical value, some codons tend to encode certain amino-acids more than the others. This signifies the emergence of the genetic code at a coding transition (dashed vertical line). Following the coding transition both the error-load and the cost increase, but this is compensated by the increasing benefit of diversity, and the overall fitness, which is their weighted sum, also increases (for explicit calculation of the coding transition see [1-4]). **(B)** The error-load is equivalent to the elastic energy of a spring framework in the shape of the codon graph, in this case the torus code (Fig. 2).

The position of each vertex $i$ is a vector $\mathbf{w}(i) = (p_{i\alpha}, p_{i\beta}, p_{i\gamma}\ldots)$, whose coordinates are the probabilities that $i$ encodes a given amino-acid (i.e., the $i$-th row of the code-matrix) and each axis corresponds to one of the amino-acids, $\alpha, \beta, \gamma\ldots$. By conservation of probability, all the vertices reside in the triangle $\Sigma_\alpha\, p_{i\alpha} = 1$ (gray). A "spring" connecting codons that encode chemically different amino-acids is stretched and costs "elastic energy", i.e., error-load. But if these codons encode the same amino-acid, then the "spring" is loose and costs no error-load. The emergence of the genetic code is manifested as the inflation of the codon-graph. Below the coding transition the coding matrix is uniform and the codon-graph collapses to a point (red dot). Above the transition, the graph starts to expand. This costs error-load but at the same time increases the diversity of the code. Eventually, the code approaches the corners of the triangle, where each codon encodes solely one amino-acid.

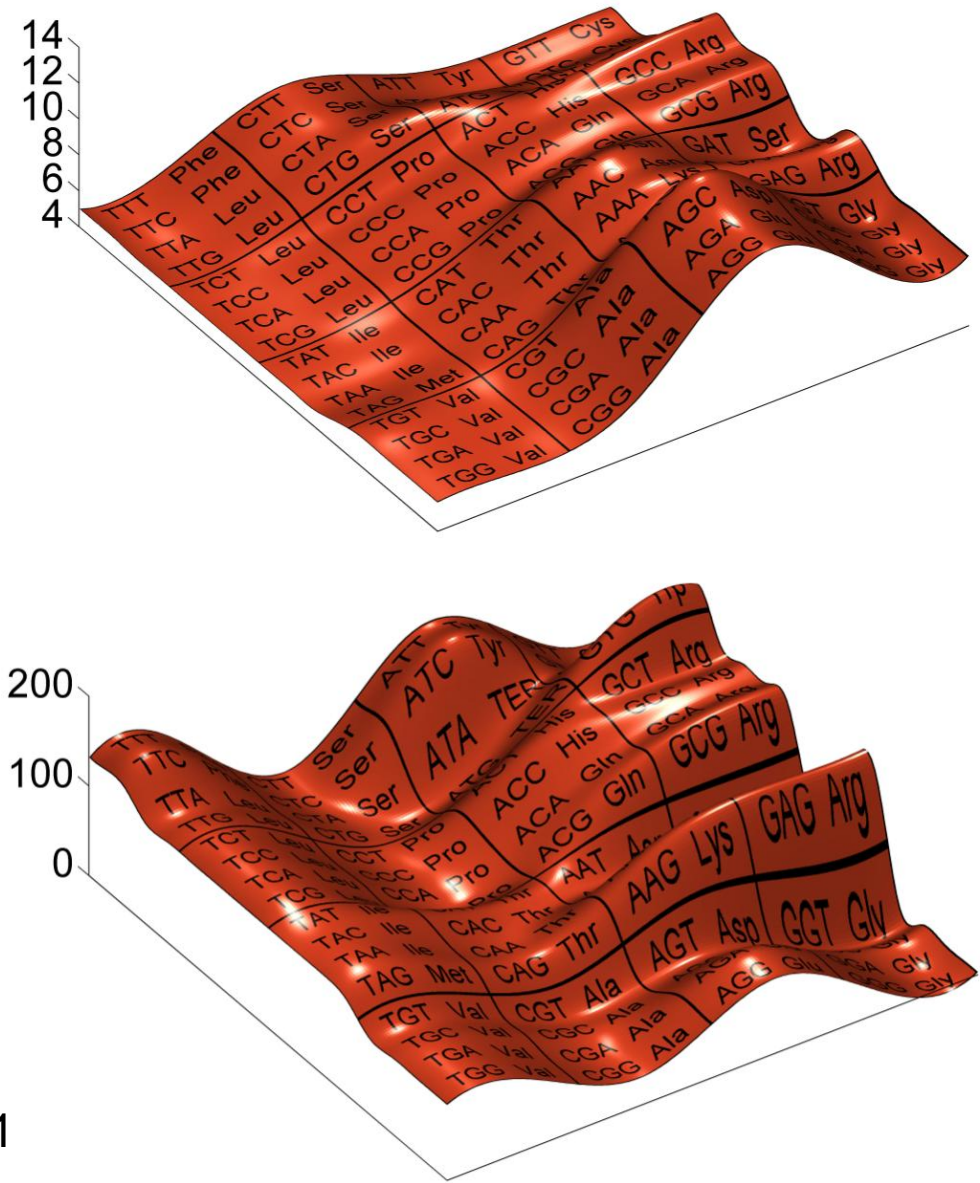

Fig. 1

**A**

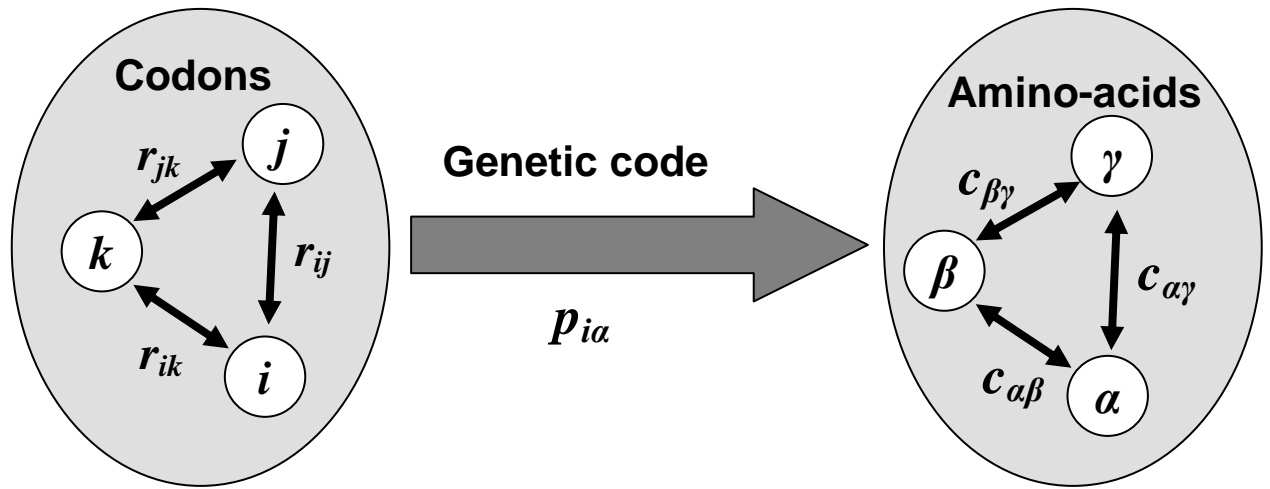

**B**

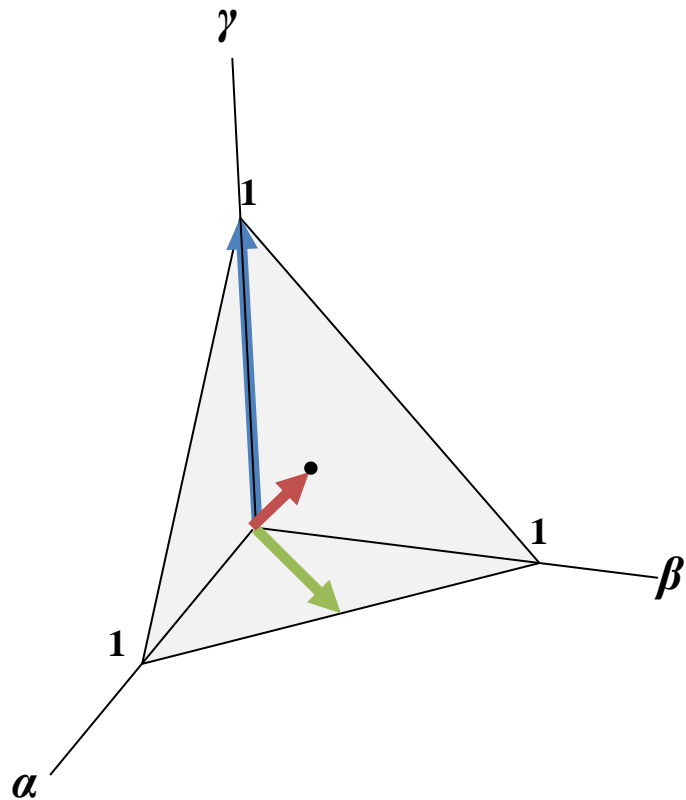

Fig. 2

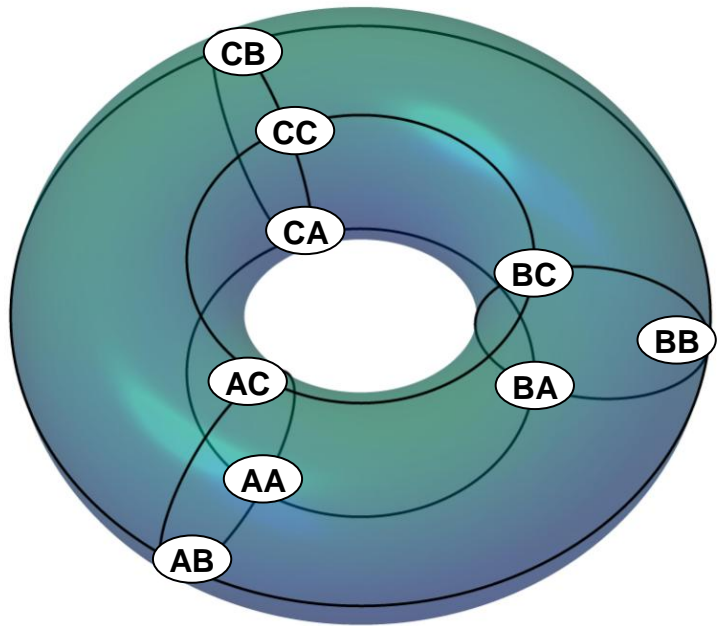

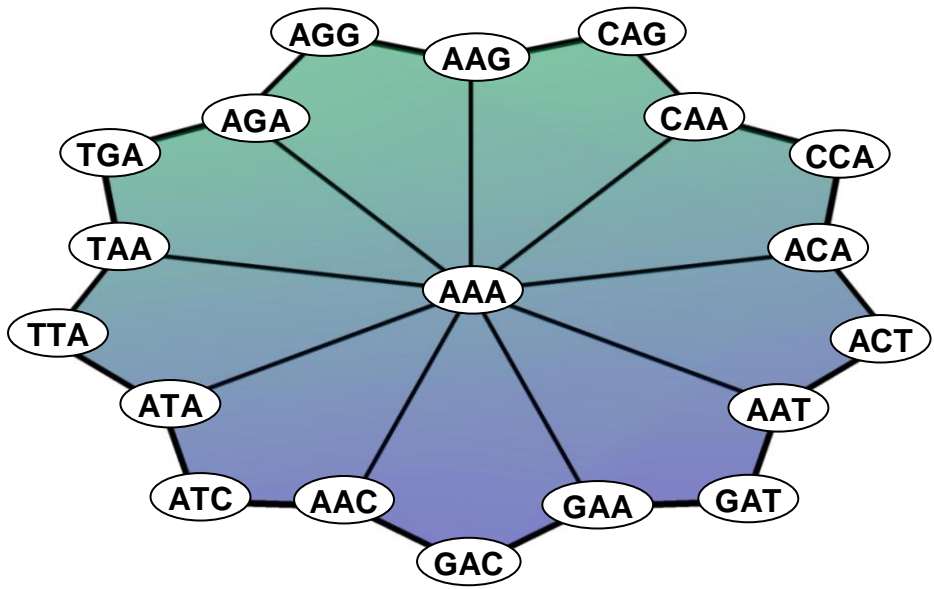

Fig. 3

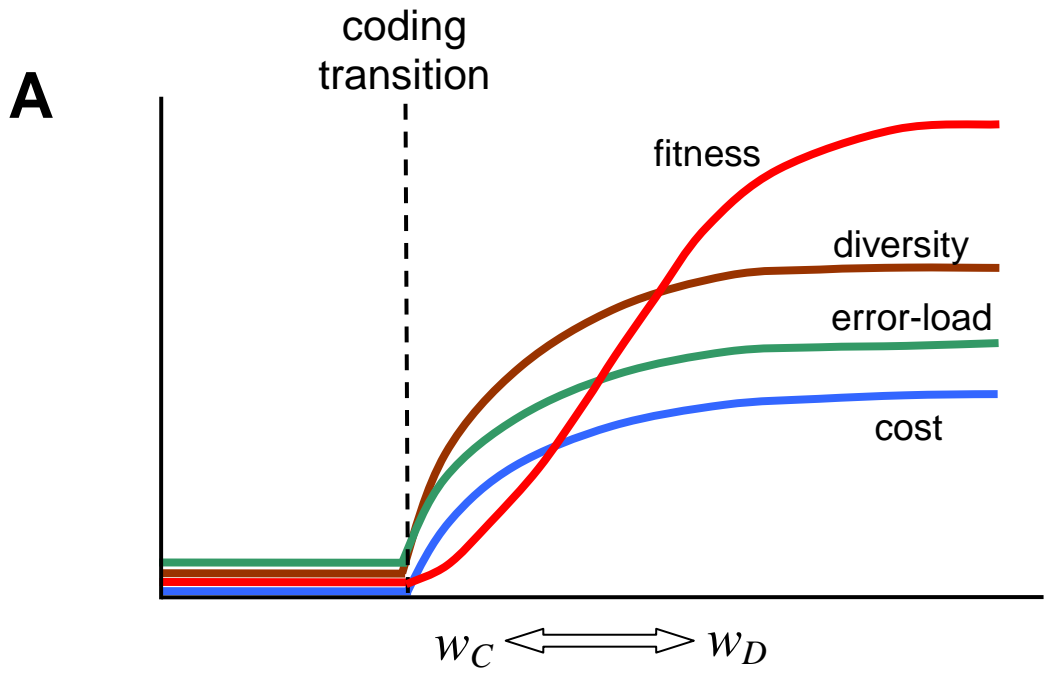

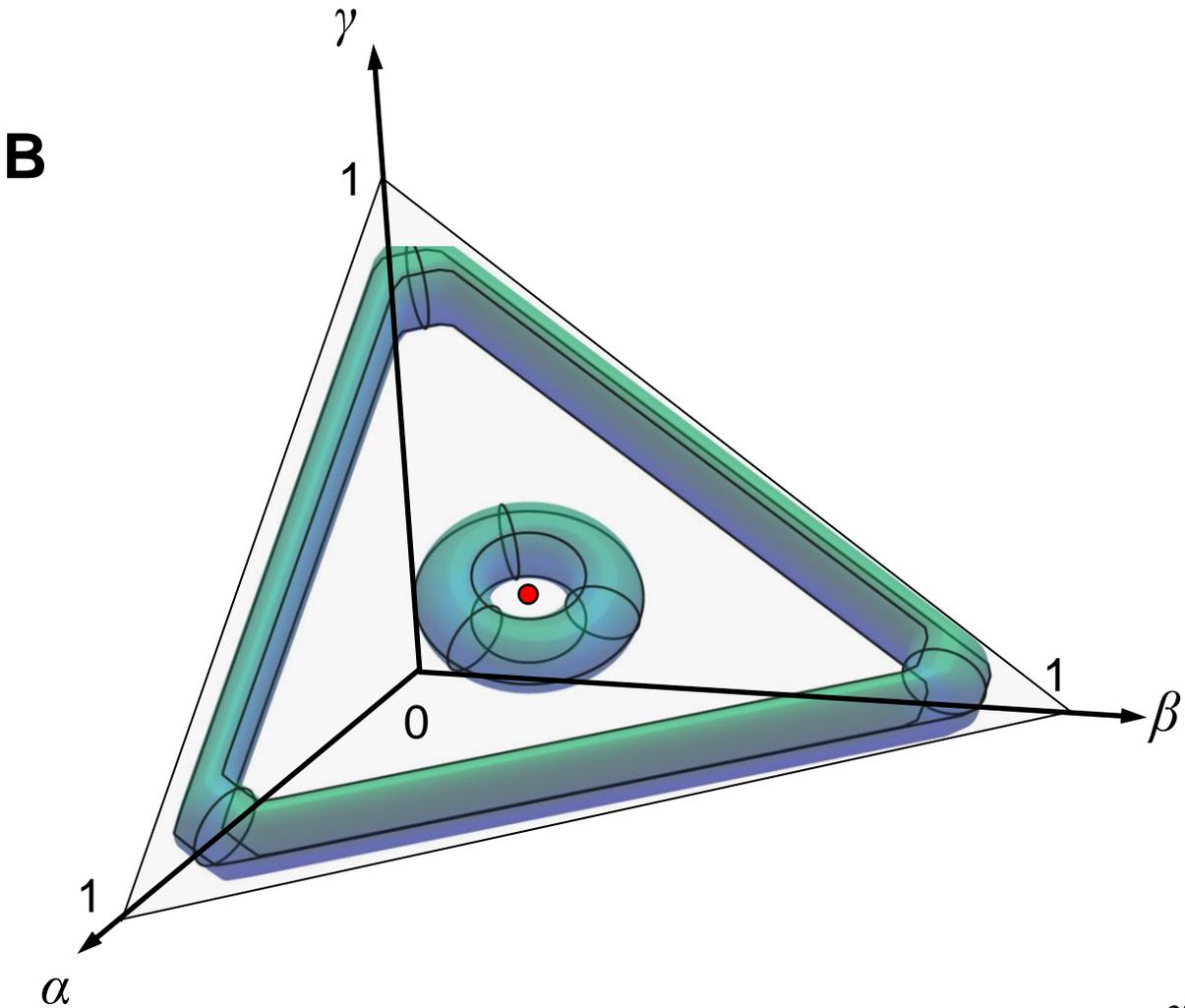

Fig. 4